# Nuclear ground-state properties from mean-field calculations

J. Dobaczewski[1], W. Nazarewicz[1−3], and M.V. Stoitsov[2−4]

[1] Institute of Theoretical Physics, Warsaw University, Hoża 69, PL-00-681 Warsaw, Poland
[2] Department of Physics & Astronomy, University of Tennessee, Knoxville, Tennessee 37996, USA
[3] Physics Division, Oak Ridge National Laboratory, Oak Ridge, Tennessee 37831, USA
[4] Institute of Nuclear Research and Nuclear Energy, Bulgarian Academy of Sciences, Sofia-1784, Bulgaria

**Abstract.** The volume and surface effects in the nuclear local energy density and the volume and surface components of the pairing interaction are discussed in the context of the mean-field, Hartree-Fock-Bogoliubov description of atomic nuclei. Predictions of properties of exotic nuclei close to the particle drip lines are presented.

**PACS.** 21.60.Jz Hartree-Fock and random-phase approximations – 21.10.Dr Binding energies and masses

## 1 Introduction

The mean-field methods are very successful in describing and predicting properties of nuclei across the chart of the nuclides. This is especially true for heavy nuclei, where the bulk properties of nuclear matter dominate over the surface effects. However, when details of nuclear structure are considered, a correct description of the nuclear surface is essential. Moreover, the surface region may give us invaluable information on the nature and strength of nuclear effective interactions in channels that are inaccessible by considering infinite systems (i.e., nuclear matter). In the present report, we briefly discuss several aspects of the surface effects in the particle-hole (Sect. 2) and particle-particle channels (Sect. 3), as well as the deformation effects (Sect. 4).

## 2 Volume and surface components of the energy density

Without any detailed microscopic knowledge of the nuclear effective interactions, we can rely on general properties of saturating fermion systems to assume that the total energy $\mathcal{E}$ of a nucleus is an integral of a local energy density $\mathcal{H}(\boldsymbol{r})$. Such a conjecture is a basis for the so-called Local Density Approximation (LDA) that has been extensively used in the context of atomic and molecular physics. In nuclear physics LDA has been employed in the form of the Skyrme-HF approximation, in which the total energy $\mathcal{E}$ is given as

$$\mathcal{E} = \mathcal{E}_{\text{kin}} + \mathcal{E}_{\text{Skyrme}} + \mathcal{E}_{\text{S-O}} + \mathcal{E}_{\text{Coul}} + \mathcal{E}_{\text{pair}}, \quad (1)$$

or equivalently

$$\mathcal{E} = \int d^3\boldsymbol{r} \; \Big[ \; \mathcal{H}_{\text{kin}}(\boldsymbol{r}) + \mathcal{H}_{\text{Skyrme}}(\boldsymbol{r}) + \mathcal{H}_{\text{S-O}}(\boldsymbol{r})$$
$$+ \mathcal{H}_{\text{Coul}}(\boldsymbol{r}) + \mathcal{H}_{\text{pair}}(\boldsymbol{r}) \Big]. \quad (2)$$

The kinetic ($\mathcal{H}_{\text{kin}}$), Skyrme ($\mathcal{H}_{\text{Skyrme}}$), spin-orbit ($\mathcal{H}_{\text{S-O}}$), Coulomb ($\mathcal{H}_{\text{Coul}}$), and pairing ($\mathcal{H}_{\text{pair}}$) densities are functions of several local densities:

$$\mathcal{H}_{\text{kin}}(\boldsymbol{r}) = \frac{\hbar^2}{2m}\left(1 - \frac{1}{A}\right)\tau_0, \quad (3)$$

$$\mathcal{H}_{\text{Skyrme}}(\boldsymbol{r}) = \sum_{t=0,1}\left[C_t^\rho(\rho_0)\rho_t^2 + C_t^{\Delta\rho}\rho_t\Delta\rho_t + C_t^\tau\rho_t\tau_t\right], \quad (4)$$

$$\mathcal{H}_{\text{S-O}}(\boldsymbol{r}) = \sum_{t=0,1}\left(C_t^{\nabla J}\rho_t\boldsymbol{\nabla}\cdot\boldsymbol{J}_t\right), \quad (5)$$

$$\mathcal{H}_{\text{Coul}}(\boldsymbol{r}) = V_{\text{Coul}}(\rho_p) - \frac{3e^2}{4}\left(\frac{3}{\pi}\right)^{1/3}\rho_p^{4/3}, \quad (6)$$

$$\mathcal{H}_{\text{pair}}(\boldsymbol{r}) = \frac{1}{4}f_{\text{pair}}(\rho_0)\sum_{t=0,1}\kappa_t^2, \quad (7)$$

where, e.g., $\rho_0=\rho_n+\rho_p$ and $\rho_1=\rho_n-\rho_p$ are the isoscalar and isovector particle densities, respectively, and $\rho_n$ and $\rho_p$ are the corresponding neutron and proton densities. For complete definitions of other densities and coupling constants appearing in expressions (3)–(7), the reader is referred, e.g., to Refs. [1,2]. (In Eq. (4) we have omitted the term depending on the tensor spin-current density because below we use only the Skyrme force in which this particular term was neglected.)

In the lower panel of Fig. 1 the above five energy densities are plotted for $^{120}$Sn, together with their sum $\mathcal{H}_{\text{tot}}$.

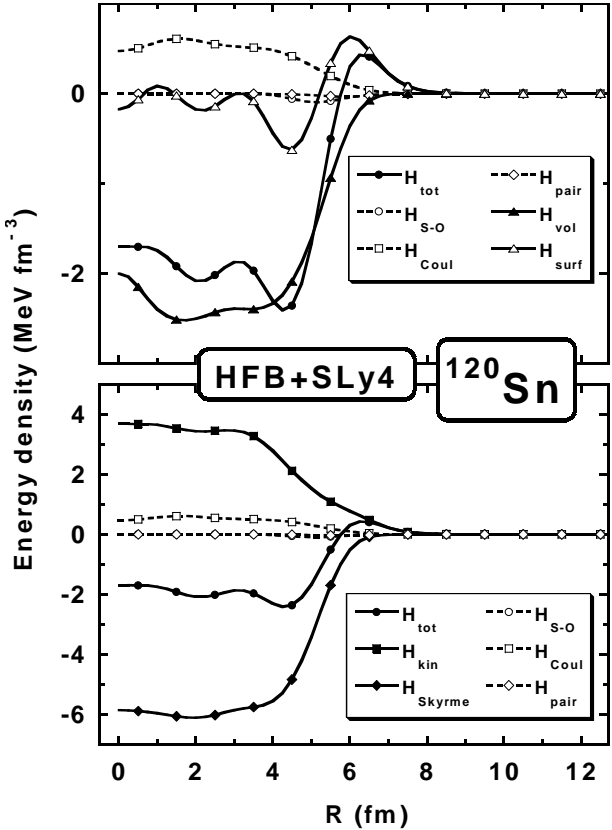

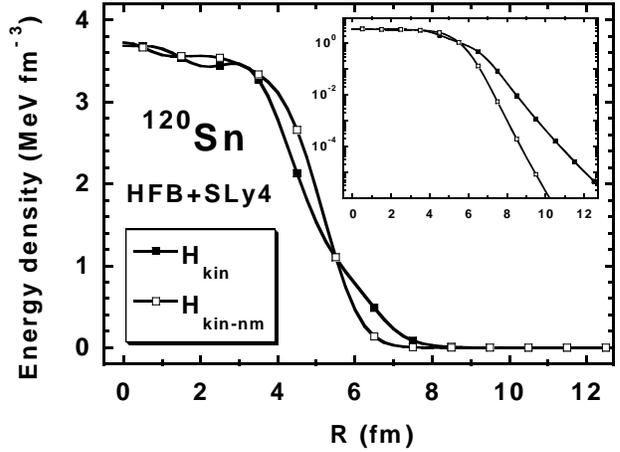

**Fig. 1.** Total energy density $\mathcal{H}_{\text{tot}}$ (full circles) calculated within the Skyrme-HF-SLy4 method for $^{120}$Sn, and plotted together with its five components given by decompositions defined in Eqs. (2) (lower panel) and (9) (upper panel).

**Fig. 2.** Microscopic kinetic energy density $\mathcal{H}_{\text{kin}}$ (full circles) compared to the nuclear-matter (volume) approximation $\mathcal{H}_{\text{kin-nm}}$ (open circles).

Calculations were performed for the Skyrme interaction SLy4 [3] and the volume pairing force. (See Ref. [4] for details of the calculations.) The presented results are very generic, and identical qualitative results are obtained for any other nucleus or interaction.

Only three out of five components significantly contribute to the total energy density, namely, the kinetic, Skyrme, and Coulomb densities. The remaining two have, of course, a decisive influence on detailed properties of nuclei, however, they are almost invisible in the scale of Fig. 1, and for the sake of the following discussion can be safely put aside. We can also see that the kinetic energy density dominates in the surface region; indeed, both the Skyrme and Coulomb terms simultaneously go to zero at a distance that is by about 1 fm smaller than the place where the kinetic energy vanishes.

Therefore, in order to analyze the energy relations at the nuclear surface, it is essential to consider surface properties of the kinetic energy density. Semiclassical methods are not appropriate to separate the volume and surface contributions to $\mathcal{H}_{\text{kin}}$, because such approaches are not valid beyond the classical turning point. Therefore, one is often fitting the volume and surface terms to reproduce the microscopic values of $\mathcal{H}_{\text{kin}}$ (see, e.g., [5]). From such analyses it turns out that the volume contribution is very

well described by the nuclear-matter expression

$$\tau_{n,p}^{\text{nm}} = \frac{\pi^{4/3}}{5}(3\rho_{n,p})^{5/3}. \tag{8}$$

Hence, in the present study we simply consider the remaining part of the kinetic energy density to be the surface contribution.

In Fig. 2 we compare the microscopic kinetic energy density $\mathcal{H}_{\text{kin}}$, Eq. (3), with the corresponding nuclear-matter (volume) contribution $\mathcal{H}_{\text{kin-nm}}$ obtained by replacing $\tau_0 = \tau_n + \tau_p$ with $\tau_0^{\text{nm}} = \tau_n^{\text{nm}} + \tau_p^{\text{nm}}$. The difference between the two curves gives our surface contribution to the kinetic energy density.

Based on these arguments, we can now rearrange terms in $\mathcal{E}$ in such a way as to single out the volume and surface contributions:

$$\mathcal{E} = \int d^3\boldsymbol{r}\ \Big[\ \mathcal{H}_{\text{vol}}(\boldsymbol{r}) + \mathcal{H}_{\text{surf}}(\boldsymbol{r}) + \mathcal{H}_{\text{S-O}}(\boldsymbol{r}) \\ + \mathcal{H}_{\text{Coul}}(\boldsymbol{r}) + \mathcal{H}_{\text{pair}}(\boldsymbol{r})\Big], \tag{9}$$

where

$$\mathcal{H}_{\text{vol}}(\boldsymbol{r}) = \frac{\hbar^2}{2m}\left(1 - \frac{1}{A}\right)\tau_0^{\text{nm}} \\ + \sum_{t=0,1}\left[C_t^\rho(\rho_0)\rho_t^2 + C_t^\tau\rho_t\tau_t^{\text{nm}}\right], \tag{10}$$

$$\mathcal{H}_{\text{surf}}(\boldsymbol{r}) = \frac{\hbar^2}{2m}\left(1 - \frac{1}{A}\right)\left(\tau_0 - \tau_0^{\text{nm}}\right) \\ + \sum_{t=0,1}\left[C_t^{\Delta\rho}\rho_t\Delta\rho_t + C_t^\tau\rho_t\left(\tau_t - \tau_t^{\text{nm}}\right)\right], \tag{11}$$

and $\mathcal{H}_{\text{vol}}(\boldsymbol{r}) + \mathcal{H}_{\text{surf}}(\boldsymbol{r}) = \mathcal{H}_{\text{kin}}(\boldsymbol{r}) + \mathcal{H}_{\text{Skyrme}}(\boldsymbol{r})$. In the Skyrme energy density (4) the effective-mass terms have been separated into the volume and surface parts according to the prescription defined above for the bare-mass terms.

In Fig. 1 (upper panel) are plotted the volume (10) and surface (11) contributions to the $^{120}$Sn total density energy. It is clear that in the surface region of this nucleus (5–7 fm) these two contributions are of a similar magnitude and opposite sign. Therefore, the nuclear surface *cannot* simply be regarded as a layer of nuclear matter at low density. In this zone the gradient terms (absent in the nuclear matter) are as important in defining the energy relations as those depending on the local density.

This observation exemplifies the difficulties in extracting the pairing properties of finite nuclei from the nuclear-matter calculations. In particular, the nuclear-matter pairing intensity, calculated at densities below the saturation point, need not be the same as the analogous intensity at the surface of a nucleus. The nuclear-matter and neutron-matter calculations of the pairing gap (see Ref. [6] for a review) performed at various densities by using very advanced and sophisticated methods, as well the best bare NN forces, can therefore be, at most, considered as weak indications of what might be the actual situation in nuclei. In this respect, calculations in semi-infinite matter that recently became available [7] may provide much more reliable information.

## 3 Volume and surface pairing interactions

Without having at our disposal microscopic first-principle effective pairing interactions (with surface effects included as discussed in Sec. 2), one uses in the particle-particle (p-p) channel a phenomenological density-dependent contact interaction. As discussed in a number of papers (see, e.g., Refs. [8,9,4]), the presence of the density dependence in the pairing channel has consequences for the spatial properties of pairing densities and fields. The commonly used density-independent contact delta interaction,

$$V_{\mathrm{vol}}^{\delta}(\boldsymbol{r}, \boldsymbol{r}') = V_0 \delta(\boldsymbol{r} - \boldsymbol{r}'), \qquad (12)$$

leads to volume pairing. A simple modification of that force is the density-dependent delta interaction (DDDI) [10,11,12]

$$V_{\mathrm{surf}}^{\delta}(\boldsymbol{r}, \boldsymbol{r}') = f_{\mathrm{pair}}(\boldsymbol{r}) \delta(\boldsymbol{r} - \boldsymbol{r}'), \qquad (13)$$

where the pairing-strength factor is

$$f_{\mathrm{pair}}(\boldsymbol{r}) = V_0 \left\{ 1 - [\rho_0(\boldsymbol{r})/\rho_c]^{\alpha} \right\} \qquad (14)$$

and $V_0$, $\rho_c$, and $\alpha$ are constants. If $\rho_c$ is chosen such that it is close to the saturation density, $\rho_c \approx \rho_0(\boldsymbol{r}=0)$, both the resulting pair density and the pairing potential are small in the nuclear interior, and the pairing field becomes surface-peaked. By varying the magnitude of the density-dependent term, the transition from volume pairing to surface pairing can be probed. A similar form of DDDI, also containing the density gradient term, has been used in Refs. [13,14].

Apart from rendering the pairing weak in the interior, the specific functional dependence on $\rho_0$ used in Eq. (14) is not motivated by any compelling theoretical arguments or calculations. In particular, values of power $\alpha$ were chosen *ad hoc* to be either equal to 1 (based on simplicity), see, e.g., Refs. [15,16], or equal to the power $\gamma$ of the Skyrme-force density dependence in the p-h channel [9,4].

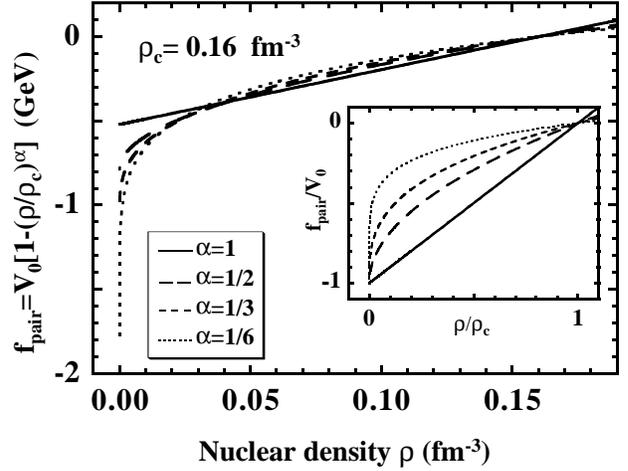

**Fig. 3.** Radial strength factor $f_{\mathrm{pair}}$ of the density-dependent delta interaction, Eq. (14), as a function of $\rho \equiv \rho_0$ for several values of $\alpha$. The value of $\rho_c$ was assumed to be 0.16 fm$^{-3}$. At each value of $\alpha$, the strength $V_0$ was adjusted to reproduce the neutron pairing gap in $^{120}$Sn. The inset shows $f_{\mathrm{pair}}/|V_0|$ as a function of dimensionless normalized density $\rho_0/\rho_c$ (from Ref. [17]).

The dependence of results on $\alpha$ was studied in Ref. [17] within the Hartree-Fock-Bogoliubov (HFB) approach. We considered four values of $\alpha=1$, $1/2$, $1/3$, and $1/6$ that cover the range of values of $\gamma$ used typically for the Skyrme forces. For $\rho_c$ we took the standard value of 0.16 fm$^{-3}$, and the strength $V_0$ of DDDI was adjusted according to the prescription given in Ref. [9], i.e., so as to obtain in each case the value of 1.245 MeV for the average neutron gap in $^{120}$Sn. The resulting pairing-strength factors (14) are shown in Fig. 3 as functions of density $\rho \equiv \rho_0$ for the four values of the exponent $\alpha$. It is seen that for $\rho \gtrsim 0.04$ fm$^{-3}$ the pairing-strength factor $f_{\mathrm{pair}}$ is almost independent of the power $\alpha$. At low densities, however, the pairing interaction becomes strongly dependent on $\alpha$ and very attractive at $\rho \to 0$. The pattern shown in Fig. 3 indicates that pairing forces characterized by small values of $\alpha$ should give rise to pair fields peaked at, or even beyond, the nuclear surface (halo region) where the nucleonic density is low.

The main conclusion of Ref. [17] is that, due to the self-consistent feedback between particle and pairing densities, the size of the neutron halo is indeed strongly influenced by pairing correlations; hence, by the pairing parametrization assumed. Consequently, experimental studies of neutron distributions in nuclei are extremely important for determining the density dependence of pairing interaction in nuclei. At the same time, the strong low-density dependence of the pairing force, simulated by taking very small values of $\alpha$ in DDDI, is unphysical. The present experimental data are consistent with about $1/2 \leq \alpha \leq 1$. In this context, it is interesting to note that excellent fits to the data were obtained in Refs. [13,14] by taking $\alpha = 2/3$. However, at present there is no theoretical argument why the density dependence should be even taken in a form of the power law.

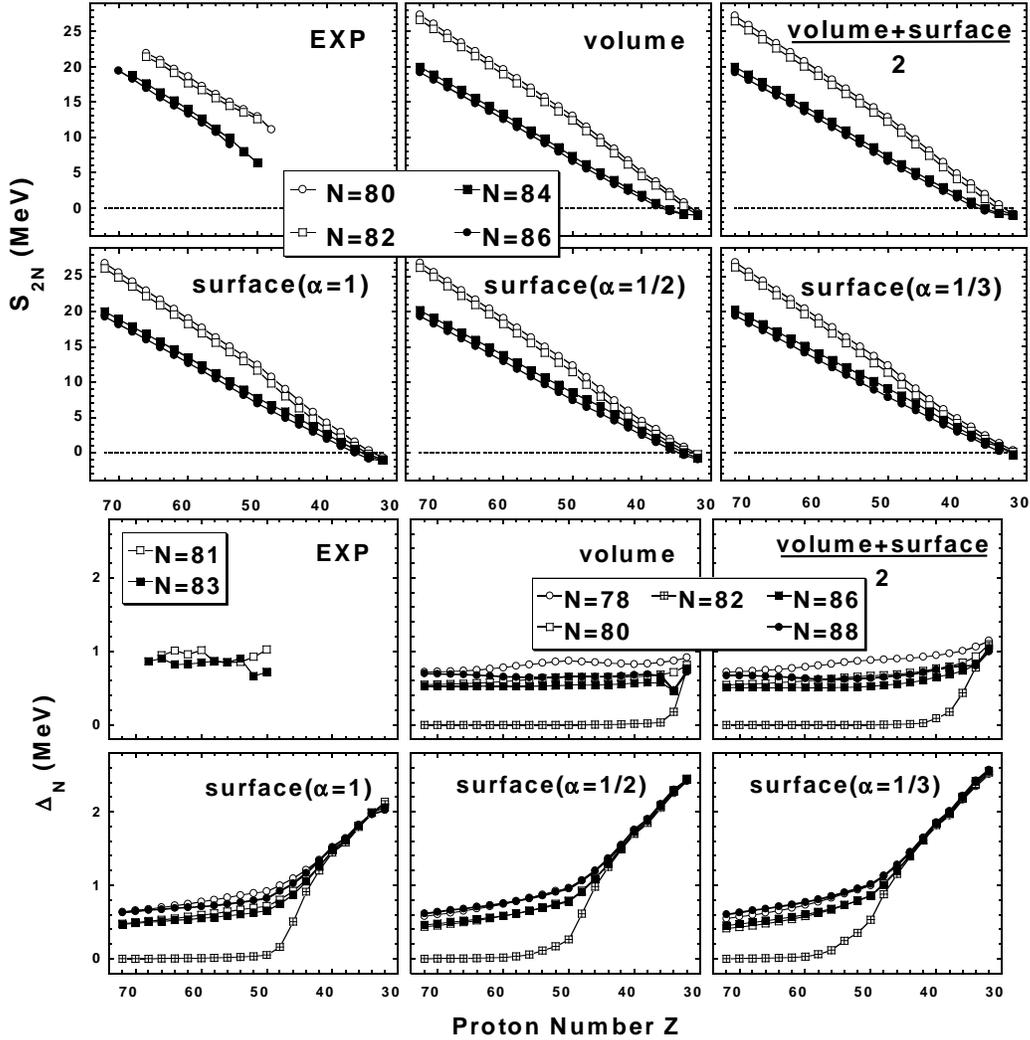

**Fig. 4.** Comparison between the experimental two-neutron separation energies $S_{2N}$ and neutron pairing gaps $\Delta_N$ (upper left panels, based on masses from Ref. [18]), and the corresponding results of the spherical HFB method for the Skyrme SLy4 force [3] and five different versions of the zero-range pairing interaction (see text).

Moreover, the pairing interaction is most likely of an intermediate character between the volume (12) and surface forms (13). (See Refs. [7,19,20] for recent analyses.) In particular, the force which is a fifty-fifty mixture of both types,

$$V_{\mathrm{mix}}^{\delta}(\mathbf{r}, \mathbf{r}') = \frac{1}{2}\left(V_{\mathrm{vol}}^{\delta} + V_{\mathrm{surf}}^{\delta}\right) = V_0 \left[1 - \frac{\rho(\mathbf{r})}{2\rho_0}\right] \delta(\mathbf{r} - \mathbf{r}'),$$
(15)

performs quite well [20] in reproducing the general mass-dependence of the odd-even mass staggering parameter $\Delta^{(3)}$ centered at odd particle numbers [21,22].

Figure 4 illustrates the role of using different types of the pairing interaction to predict the two-neutron sep-aration energies and neutron pairing gaps, respectively, in very neutron-rich isotones around $N{=}82$. The experimental values were calculated based on the interim 2001 evaluation of atomic masses [18].

Figure 4 nicely illustrates the effect of the so-called shell quenching in heavy nuclei [23], i.e., the vanishing of the effective distance between the neutron single-particle levels above and below a magic neutron number when approaching the neutron drip line. The difference between the two-neutron separation energies above and below $N = 82$ very well visualizes this effect. In fact, the experimental data show an apparent opposite effect; however, this is caused by the fact that the data are available only for $Z{\geq}50$. When approaching the magic proton number, the neutron magic gap is slightly enhanced [24]. This effect is entirely absent in calculations that do not include any effects of correlations and deformations.

Nevertheless, for $Z{<}50$ the effect of the shell quenching is very well visible in the calculations. Moreover, the magnitude of the effect is very strongly influenced by the type of pairing force used. For the volume pairing force (12), the effect is rather weak and the magic gap $N=82$ is still visible even at the very drip line. However, for the surface pairing force (13) the shell gap goes to zero much earlier, and this tendency is accentuated for pairing forces that are stronger at small densities (for smaller powers of $\alpha$).

For the neutron pairing gap (Fig. 4) the experimental data that exist for $Z \geq 50$ do not indicate any definite change in the neutron pairing intensity with varying proton numbers. However, the surface pairing interactions (bottom panels) give a slow dependence for $Z \geq 50$ that is dramatically accelerated after crossing the shell gap at $Z=50$. On the other hand, the volume and intermediate-type pairing forces predict a slow dependence all the way through to very near the neutron drip line. It is clear that measurements of only several nuclear masses for $Z<50$ will allow us to strongly discriminate between the pairing interactions that have different space and density dependencies.

## 4 Deformation of drip-line nuclei

Deformation of nuclei near the drip lines is a difficult and open problem in nuclear structure physics. It requires a simultaneous description of particle-hole, pairing, and continuum effects — the challenge that only very recently can be addressed by mean-field methods. Deformability of nuclei plays a decisive role in determining particle separation energies and decay rates, and hence is crucial for a description of nuclear processes in a stellar environment.

Very recently we have developed methods [25,26] to approach this problem by using the local-scaling point transformation that allows us to modify asymptotic properties of the deformed harmonic oscillator wave functions. The resulting single-particle bases are very well suited for solving the HFB equations for deformed drip-line nuclei.

Calculations of a complete HFB mass table are now in progress and will be reported in separate publications [27, 28]. Here we only show a sample of the results obtained for the SLy4 Skyrme interaction and the intermediate-type pairing force (15). A rather restricted size of the harmonic oscillator basis, limited to $N_{\rm sh}=14$ spherical shells, was used, while the continuum states were included up to 60 MeV. Figure 5 shows the two-neutron separation energies (top panel), the neutron pairing gaps (middle panel), and the deformations $\beta$ (bottom panel) for 1553 particle-bound even-even nuclei with $Z \leq 108$ and $N \leq 188$.

At the two-neutron drip line, one can see a very interesting effect of negative two-neutron separation energies for particle-bound (negative Fermi energy) nuclei. This is the result of a sudden change in configuration when approaching the drip line, that is caused by the fact that the ground-state configuration may become particle-unbound earlier than the excited one [25,27]. A similar effect occurs also in the heavy proton drip-line nuclei, where sequences of oblate ground states are obtained. Another effect at the proton drip line is related to long sequences of proton-magic (e.g., $Z=50$ and 82) isotopes intruding in the territory of unbound nuclei. This is the result of the vanishing pairing correlations, for which the proton Fermi energy coincides with the last occupied level, while in the neighboring nuclei it is located higher.

This research was supported in part by the U.S. Department of Energy under Contract Nos. DE-FG02-96ER40963 (University of Tennessee) and DE-AC05-00OR22725 with UT-Battelle, LLC (Oak Ridge National Laboratory), by the Polish Committee for Scientific Research (KBN) under Contract No. 5 P03B 014 21, and by computational grants from the *Regionales Hochschulrechenzentrum Kaiserslautern* (RHRK) Germany, European Center for Theoretical Studies (ECT*) in Trento, Italy, and Interdisciplinary Centre for Mathematical and Computational Modeling (ICM) of Warsaw University.

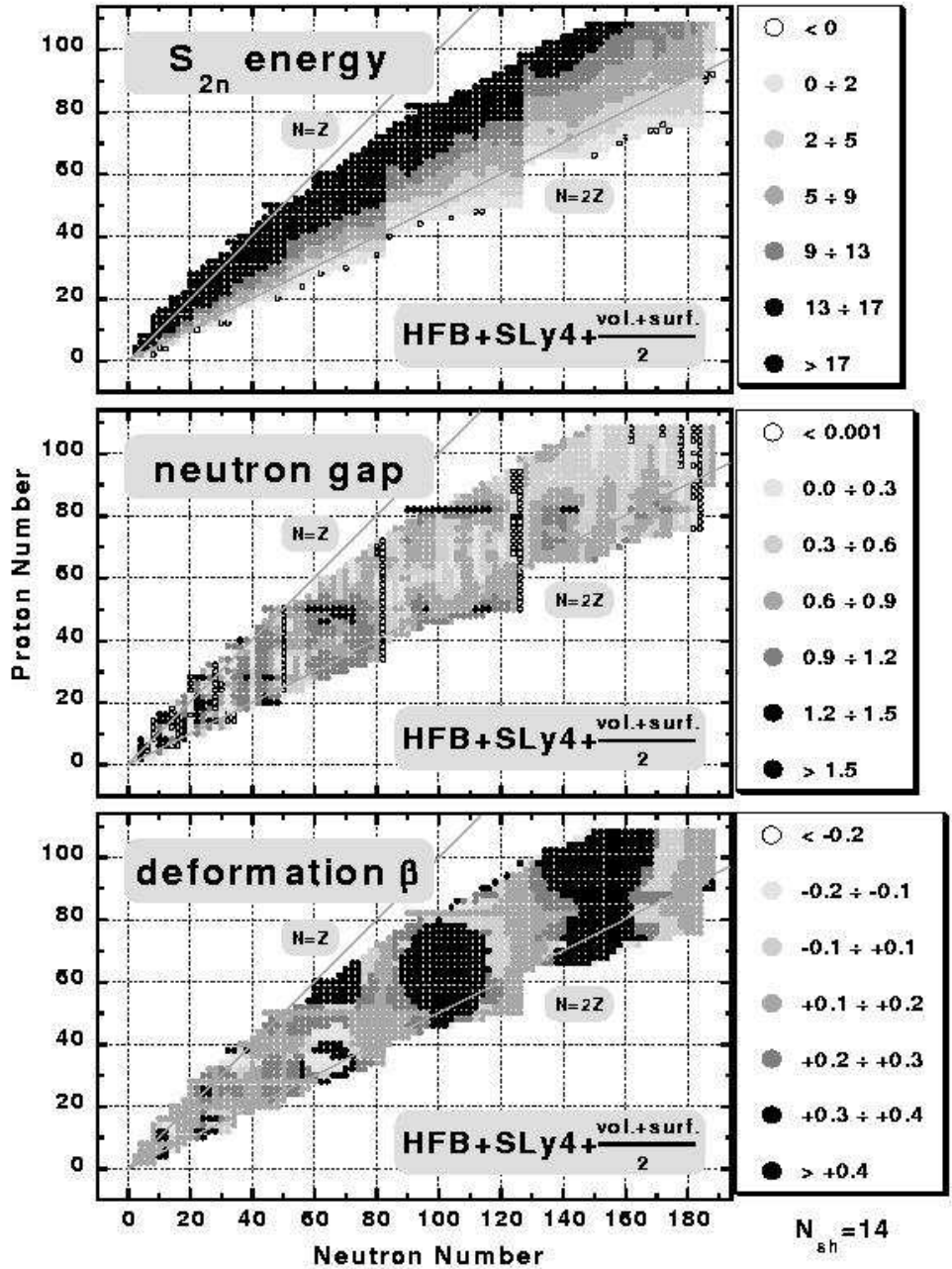

**Fig. 5.** Results of the HFB+THO+SLy4 deformed calculations for particle-bound even-even nuclei with $Z \leq 108$ and $N \leq 188$.